\documentclass[12pt,preprint]{aastex}

\slugcomment{}

\shorttitle{The Total luminosity from Quasars }
\shortauthors{Elvis, Risaliti \& Zamorani}

\begin{document}
\title{Most Supermassive Black Holes must be Rapidly Rotating}

\author{M. Elvis\altaffilmark{1}, G. Risaliti\altaffilmark{1,2}
\& G. Zamorani\altaffilmark{3}}

\altaffiltext{1}{Harvard-Smithsonian Center for Astrophysics, 60
Garden Street, Cambridge, MA 02138}
\altaffiltext{2}{Osservatorio Astrofisico di Arcetri,
Largo E. Fermi 5, I-50125 Firenze, Italy}
\altaffiltext{3}{Osservatorio Astronomico di
Bologna, via Ranzani 1, I-40127 Bologna, Italy}

\begin{abstract}
We use the integrated spectrum of the X-ray background and quasars Spectral
Energy Distribution to derive the contribution of quasars to the energy output
of the Universe. We find a lower limit for  the energy from accretion onto 
black holes of 6\%,
of the total luminosity of the Universe and probably more,
with 15\% quite possible. Comparing these values with the masses of
black holes in the center of nearby galaxies we show that the accretion process
must be on average very efficient: at least 15\% of the accreted mass must
be transformed into radiated energy. This further implies that most supermassive black holes
are rapidly rotating.
\end{abstract}

\keywords{galaxies: active --- X-rays: diffuse background}
\section{Introduction}

 
The recent realization that most of quasars are hidden behind
large amounts of dust and gas (Fabian \& Iwasawa 1999, hereafter FI99,
Maiolino \& Rieke 1995) has increased the estimate of the total
quasar luminosity of the universe substantially, to a point where
it is no longer negligible compared with that of stars, and to a
level where it fully explains the diffuse background seen in X-rays.
Since Hubble observations have determined the mean mass density of
black holes, the ratio of luminosity to mass determines a maximum
efficiency required. If this substantially exceeds 5\% then
supermassive black holes must, on average, be rapidly spinning.
 
FI99 demonstrated that assuming 10\%
efficiency the X-ray background and the black hole mass density
were consistent. Here we revisit their argument, carefully examining
all the assumptions, to produce a strong lower limit on the
supermassive black hole efficiency. We find that this efficiency
must be greater than 15\%.
Most supermassive black holes must then be spinning.

\section{The total luminosity from quasars}

The hard X-ray background spectrum (XRB, E$>$ 2 keV) is accurately reproduced
by the integrated emission of AGNs, most of
which have to be obscured, (Setti \& Woltjer 1989, Comastri et al. 1995,
Gilli et al. 2001).
These synthesis models successfully predict the Chandra and XMM-Newton hard
X-ray source counts (Hasinger et al. 2001). With this increasingly well supported model
the X-ray background spectrum can be used, together with a quasar 
SED, to estimate the total energy output of AGNs in the Universe, as shown by FI99.

A key step in the argument of FI99 is the bolometric correction assumed to derive the total quasar
luminosity from the X-ray intensity. The SED assumed by FI99  is that of Elvis et al. (1994, hereafter E94)),
obtained
from a sample of quasars for which an Einstein IPC spectrum was available.
Here 1) we derive a different bolometric correction by taking into account the statistical bias on the average
X-ray to optical ratio derived by the analysis of an X-ray selected sample and by excluding the infrared
region (which would otherwise double count the UV and soft X-ray photons); 2) we discuss the possibility
that this ratio decreases with redshift; and 3) we complete the Elvis et al. SED with the recent UV SED
of Cheng et al. 1999. All together, these changes increase by a factor $>$ 2 the estimate of the quasars'
total energy output with respect to the value derived by FI99.

We can estimate the total energy released by quasars with the following two steps:
(1) we assume that the X-ray background at 30 keV is due to unobscured emission from AGNs and
then extrapolate the value of the {\em unobscured} background down to 2 keV, $\Psi_{2 keV}=E\times
I(E) (2 {\rm keV})$,
assuming an average
X-ray spectrum for quasars; (2) we compute the total energy output from quasars,$\Psi_T$, using an average
quasar SED. The first step involves a number of possible corrections, each of which of the order of 30\%, with respect
to the simple case of a single power law slope adopted by FI99; the second step leads to a bolometric correction
of a factor 1.6-3 higher than that of FI99.

1)  The spectral complexity
of quasars in the X-rays leads  to four small uncertainties in estimating
the unobscured emitted 2 keV X-ray background:\\
(a) {\em Photon index}:
the value $\Gamma=2$ used by FI99 comes from the 
subsamples of low redshift optically selected PG quasars (Schmidt \& Green 1983,
George et al. 2000, Mineo et al. 2000).
However, Reeves et al. (2000)  find
$\Gamma=1.89\pm 0.05$ for all radio quiet quasars observed by ASCA (including several
high redshift objects), which would decrease $\Psi_{2~keV}$ by 25\%. and may be due
to a flattening of the average X-ray spectrum of quasars at high redshifts. We note though that since
this sample is X-ray selected there could be a bias
towards hard X-ray loud objects.\\
(b) {\em Reflection component}:
a reflection component in quasars spectra could decrease
the prediction of $\Psi_{2 keV}$. However, the effect should be small ($<$ 5\%), especially since
the reflection component is weaker in high luminosity quasars (Mineo et al. 2000).\\
(c) {\em High energy cutoff}:
BeppoSAX observations (Guainazzi et al. 1999, Perola et al. 1999, Perola et al. 2000, Nicastro et al. 2000,
Risaliti et al. 2001) 
find  exponential cutoffs in AGN spectra at
energies E$>100$ keV. Assuming an average redshift of 2 for quasars
making most of the XRB, and a cut off energy E=300 keV, the normalization of the continuum powerlaw (and
so of $\Psi_{2~keV}$) is increased by 35\%.\\
(d) {\em Extremely High N$_H$ objects}:
quasars absorbed by column densities N$_H > 10^{24}$ cm$^{-2}$ can be significantly
absorbed even at a rest frame energy corresponding to 30 keV in the observer frame. Moreover, the
X-ray emission is completely absorbed if  N$_H > 10^{25}$ cm$^{-2}$. According to Risaliti et al. (1999)
objects with  N$_H > 10^{25}$ cm$^{-2}$ account for at least 20\% of low luminosity AGNs at low redshift,
and could be more
than 30\%, while another 10-20\% have 10$^{24}$ cm$^{-2} <$ N$_H < 10^{25}$ cm$^{-2}$. Therefore,
if these percentages remain the same at all luminosities, we must increase $\Psi_{2~keV}$ by {\em at least} 20\%.

To estimate $\Psi_{2 keV}$ we used the code of Gilli et al.
(2001), which reproduces the XRB spectrum and source counts (Akiyama et al. 1999, Fiore et al. 1999, Hasinger et
al. 2001). 
This code uses the soft X-ray luminosity function of AGNs and its evolution with redshift (Miyaji, Hasinger \& Schmidt
2000), 
and assumes an average X-ray SED for type 1 (unobscured) and type 2 (obscured) AGNs.
To estimate the total intrinsic luminosity emitted by AGNs,
we assumed an unabsorbed SED for all AGNs with
$\Gamma=1.9$, E$_C$=300 keV. These parameters
predict a relatively low XRB intensity at 2 keV: a much higher value is quite possible, given the parameter space
allowed by the observational results described above.
The result we obtain is $\Psi_{2~keV}=40$ keV s$^{-1}$ cm$^{-2}$ sr$^{-1}$ (only 5\% larger than
38 keV s$^{-1}$ cm$^{-2}$ sr$^{-1}$, of FI99). Assuming a fraction
of completely obscured objects $f_{\rm abs}$=0.2 we obtain $\Psi_{2~keV}=48$ keV s$^{-1}$ cm$^{-2}$ sr$^{-1}$.
The Gilli et al. model predicts an increase
of the fraction of absorbed objects with redshift, so this value, already a lower limit for low redshift AGNs,
is even more conservative at high redshifts.

2) We now use $\Psi_{2~keV}$, together with a
 bolometric correction, to estimate the total
integrated emission of quasars, $\Psi_T$. At wavelengths from the radio to the UV we
can assume that the SED by E94 is representative for the average emission of unobscured
quasars. However, the X-ray emission of quasars in the E94
sample included only those quasars that gave a large enough number of photons in the Einstein IPC
to construct a well-constrained spectrum. Since the exposure times for the optically selected
 quasars were all similar, the E94 sample is biased toward X-ray brighter quasars. This is a sizeable effect.
The commonly used indicator for the X-ray
to optical ratio in quasars is the $\alpha(ox)$ index, defined as $\alpha(ox)=-\log(f_x/f_o)/2.605$,
where $f_o$ and $f_x$ are the monochromatic fluxes at 2500\AA~and 2 keV.

Given a sample of AGNs with an observed normalized distribution $D(\alpha(ox))$ of the $\alpha(ox)$ parameter,
we can define an effective $\alpha(ox)$ (see Avni et al. (1980) for details) as:
\begin{equation}
<f_o/f_x> = 10^{2.605\times \alpha(ox)_{eff}}= \int  D(\alpha(ox) )\times 10^{2.605\times \alpha(ox)}d \alpha(ox)
\end{equation}

Since our starting point is an integrated X-ray emission (i.e. $\Psi(X)$), the corresponding $<f_o/f_x>$ to be used for the
estimate of the appropriate X-ray to optical conversion is obtained from Eq. 1 with the parameters typical
of X-ray selected samples. Assuming, consistently with the data, a Gaussian distribution of $\alpha(ox)$, from Eq. 1
we obtain 
\begin{equation}
\alpha(ox)_{eff} = <\alpha(ox)>+3\sigma^2
\end{equation}
where $<\alpha(ox)>$ and $\sigma^2$ are the mean and variance of $D(\alpha(ox)$).
A reasonable value for $\sigma$ is $\sim 0.16$, in agreement with Yuan et al. 1998.
Using $<\alpha(ox)>=1.35$ (from E94, and in agreement with the results from the ROSAT deep
surveys (Lehmann et al. 2001), we obtain  $\alpha(ox)_{eff}=1.43$.

Reassuringly, we obtain the same result starting from the optically selected sample of PG quasars
(Schmidt \& Green 1983). In this case we have  $<\alpha(ox)> = 1.55$ for the subsample of low redshift
PG quasars (Laor et al. 1997). If $\alpha(ox)$ has a gaussian distribution, the relation between the observed
values of $\alpha(ox)$ in flux limited X-ray and optically selected samples (assuming the same parent
distribution) is
\begin{equation}
<\alpha(ox)>_{opt} -  <\alpha(ox)>_x  = 6\sigma^2 (\beta-1)
\end{equation}
where $\beta$ is the differential slope of the
counts as a  function of  flux (see Francis et al. (1993) for a similar discussion in a different context). Assuming the Euclidean
slope $\beta=2.5$, the observed $<\alpha(ox)>$ for this optically selected sample is approximately equivalent
to our adopted $<\alpha(ox)>$ for an X-ray selected sample and therefore consistent with $\alpha(ox)_{eff}=1.43$.
The use of this value increases
the bolometric correction by a factor 1.6, with respect to the value $\alpha(ox)=$1.35 used by FI99.

Most of the XRB originates at substantial redshifts, z$\sim 1-2$.
A study of the optically selected quasars observed by ROSAT (Yuan et al. 1998) suggests a possible increase
of  $\alpha(ox)$ with redshift or luminosity to an average $\alpha(ox)$=1.68  at z=2. If this increase
is due to a luminosity effect, since most of the XRB is made by objects of moderate luminosity (L(2-10 keV)$<10^{45}$
erg s$^{-1}$),
the safest assumption is to assume the  $<\alpha(ox)>$ value obtained
with the complete low redshift sample, which is free from biases and samples well the relevant luminosity range.
If, instead, the increase in $<\alpha(ox)>$ is mainly correlated with redshift, then it would affect the bolometric correction.
We can estimate an upper limit to this correction by using the value $<\alpha(ox)> = 1.68$ at z=2 (Yuan et al. 1998).
The corresponding $<\alpha(ox)>_{eff}$ would be $\sim 1.55$ and the bolometric
correction would be 3.3 times higher than that estimated by FI99.

3) Final source of uncertainty is the EUV SED of quasars. The SED of E94 does not cover the energy
range between the H edge (E=13.6 eV) and E$\sim$ 0.15 keV. Here we use the UV SED obtained by
Zheng et al. (1999) from a large sample of radio quiet quasars observed with HST in the 350\AA-912\AA~range.
From 350\AA~to 0.15 keV we assume a power law spectrum (the details of the SED in this band affect the final
bolometric correction by less than 5\%).
The choice of this UV SED is, once again, conservative in the sense to predict the lowest possible bolometric
correction. For comparison, the SED estimated by Mathews \& Ferland (1987), based on the equivalent width of
high ionization emission lines, leads to a $\sim 25$\% higher bolometric correction.

Using the lower limit at 2 keV computed above and assuming  $\alpha(ox)_{eff}$
to be in the range of 1.43-1.55 we obtain the integrated emission of AGNs:
$\Psi_{T} \geq (3.6-8) {\rm~ nW~m}^{-2}~{\rm sr}^{-1}$, where the first value is likely
to be a lower limit. This estimate has to be compared with the value of
$\sim 2 {\rm~ nW~m}^{-2} {\rm sr}^{-1}$, obtained by FI99
(actually FI99 obtain a value of 3 in the same units,
but, as they say in the Discussion, this value has to be lowered by 1/3, in order not to count the UV emission
twice, since the IR radiation is due to reprocessing of UV emission).

Converted into an energy density this gives:
\begin{equation}
U_T = \frac{4 \pi \Psi}{c} \sim (1.5-3.4) \times 10^{-15} {\rm ~erg~ cm}^{-3} 
\end{equation}
If we assume that this energy comes from accretion with efficiency $\epsilon$, the mass density in
black holes is:
\begin{equation}
\rho_\bullet =  \frac{1+z}{\epsilon} ~\frac{U_{T}}{c^2} \sim (7.5-16.8) \times 10^4 ~\epsilon^{-1}
M_\odot {\rm  ~Mpc}^{-3}
\end{equation}
where we assume that the peak of AGN emission is at z=2, in agreement with the XRB synthesis models.

Our results can be compared with estimates of the total luminosity in the Universe from both stars and
supermassive central black holes in galaxies.

The total luminosity of the Universe has been estimated
from the diffuse background from sub-mm to UV wavelengths. Madau \& Pozzetti (2000) estimate that
the diffuse background between 0.1 and 7 $\mu$m has an integrated intensity
$\Psi_{UO} = 17\pm 3$ nW m$^{-2}$
sr$^{-1}$. The bolometric emission in the far infrared, $\Psi_{IR}$,
between 1000$\mu$m and 7$\mu$m, derived from the IR background,
is $\sim 40$ nW m$^{-2}$ sr$^{-1}$ (Franceschini et al. 2001).
The stellar produced background outside these spectral regions is negligible.
Therefore, the total energy in the background, due to stars and quasars combined, is $\Psi_{TOT} \sim 57$
nW m$^{-2}$ sr$^{-1}$.
This means that quasars contribute {\em at least} 6\% to the total luminosity of the Universe.
This is a lower limit only dependent  on our knowledge of the SED of quasars. Since we obtained
our estimate through a chain of conservative assumptions,
a significantly higher contribution is highly probable. A value $\sim 15$\% could apply if
the redshift evolution of the optical to X-ray ratio is real, 
and up to 20-25\% if we release our conservative assumptions
on quasar spectra and the fraction of Compton thick objects (for example, a photon index $\Gamma=2$
instead of the adopted $\Gamma=1.9$,
and a fraction of quasars with N$_H>10^{25}$ cm$^{-2}$ f=30\%
instead of f=20\% are quite possible, if not favored by the hard X-ray data now available).

\section{Accretion efficiency in AGNs}
 The mass density in central black holes has been estimated by
Salucci et al. (1999) to be $\rho_\bullet \sim 4 \times 10^5 ~M_\odot ~ {\rm Mpc}^{-3}$ using a quasar evolution
model and starting from X-ray counts. These results are in agreement with a recent work (Merritt \& Ferrarese 2001) 
on the relation between central black hole mass and the velocity dispersion
of the bulge (best estimate  $\rho_\bullet \sim 5 \times 10^5 ~M_\odot ~ {\rm Mpc}^{-3}$)

Combining these estimates with
Equation 1 gives a value for the average
accretion efficiency, $\epsilon$.
We find $\epsilon > 0.15$.

Averaged over the history of the Universe our result implies that every black hole is radiating
at $>15$\% efficiency. This indicates that the accretion
history of the Universe cannot be on average silent, and rules out the possibility that a significant fraction of the
black hole mass accretion has occurred in the ADAF regime. Moreover, this value is well above
the maximum efficiency of non-rotating black holes, ($\epsilon_{NR}=5.6$\%).
Therefore, most supermassive black holes must be
rapidly spinning.\\

\acknowledgments

This work was supported in part by NASA grant NAG5-4808.


\begin{thebibliography}{}
\bibitem[]{a10} Akiyama, M., et al. 2000, ApJ, 532, 700
\bibitem[]{a15} Avni, Y., Soltan, A., Tananbaum, H., \&
                Zamorani, G. 1980, ApJ, 238, 800
\bibitem[]{a20} Comastri, A., Setti, G., Hasinger, G., Zamorani,
                G., 1995, A\&A, 296, 1
\bibitem[]{a30} Elvis, M., et al., 1994, ApJS, 95, 1  (E94)
\bibitem[]{a40} Fabian, A.C., Iwasawa, K., 1999, MNRAS 303, L34
\bibitem[]{a43} Fiore, F., La Franca, F., Giommi, P., Elvis, M.,
                Matt, G., Comastri, A., Molendi, S., \& Gioia, I. 1999, MNRAS,
                307, L55
\bibitem[]{a50} Franceschini, A., Aussel, H., Cesarsky, C. J., Elbaz, D.,
\& Fadda, D., 2001, A\&A, 378, 1
\bibitem[]{a51}	Francis, P. 1993, ApJ, 407, 519
\bibitem[]{a55} George, I., Turner, T. J., Yaqoob, T., Netzer,
                H., Laor, A., Mushotzky, R. F., Nandra, K., \& Takahashi,
                T. 2000, ApJ, 531, 52
\bibitem[]{a65} Gilli, R., Hasinger, G., \& Salvati, M., 2001,
                A\&A, 366, 407
\bibitem[]{a67} Guainazzi, M., et al. 1999, A\&A, 346, 407
\bibitem[]{a71} Hasinger, G., et al. 2001, A\&A, 365, L45
\bibitem[]{a74} Laor, A., Fiore, F., Elvis, M., Wilkes, B., \& McDowell, J. C.
                1997, ApJ, 477, 93
\bibitem[]{a77} Lehmann, I., et al. 2001, A\&A, 371, 833
\bibitem[]{a81} Nicastro, F., et al. 2000, ApJ, 536, 718
\bibitem[]{a85} Madau, P., \& Pozzetti, L. 2000, MNRAS, 312, L9
\bibitem[]{a89} Macchetto, D., Marconi, A., Axon, D. J., Capetti, A., Sparks, W.,
                \& Crane, P. 1997, ApJ, 489, 579
\bibitem[]{a90} Maiolino, R., Rieke, G. 1995, ApJ, 454, 95
\bibitem[]{a90} Marconi, A., et al. 2001, ApJ, 549, 915
\bibitem[]{a93} Mathews, W.G., \& Ferland, G.J. 1987, ApJ, 323, 546
\bibitem[]{a94} Merritt, D., \& Ferrarese, L. 2001, MNRAS, 320, L30
\bibitem[]{a94} Mineo, T., et al., 2000, A\&A, 359, 471
\bibitem[]{a95} Miyaji, T., Hasinger, G., \& Schmidt, M. 2000, A\&A, 353, 25
\bibitem[]{a96} Nicastro, F., et al. 2000, ApJ, 536, 718
\bibitem[]{a97} Perola, G.C., et al. 1999, A\&A, 351, 937
\bibitem[]{a98} Perola, G.C., et al. 2000, A\&A, 358, 117
\bibitem[]{a99} Reeves, N., \& Turner, M. J. L. 2000, MNRAS, 316, 234
\bibitem[]{a100} Risaliti, G., Maiolino, R., Salvati, M., 1999,
                ApJ, 522, 157
\bibitem[]{a100} Risaliti, G. 2001, A\&A, submitted
\bibitem[]{a102} Salucci, p., Szuszkiewicz, E., Monaco, P., \&
                Danese, L. 1999, MNRAS, 307, 637
\bibitem[]{a105} Setti, G., \& Woltjer, L. 1989, A\&A, 224, L21
\bibitem[]{a107} Schmidt, M., \& Green, R.F. 1983, ApJ, 269, 352
\bibitem[]{a108} Shapiro, L., Teukolsky, S. A. 1983,
                ``Black Holes, White Dwarfs, and Neutron Stars'',John Wiley \&
                Sons Ed. \\
\bibitem[]{a109} Wandel, A. 1999, ApJ, 519, 39
\bibitem[]{a110} Yuan, W., Brinkmann, W., Siebert, J., \& Voges,
                W. 1998, A\&A 330, 108
\bibitem[]{a120} Zheng, W., Kriss, G.A., Telfer, R.C., Grimes,
                J.P., \& Davidsen, A.F. 1997, ApJ, 475, 469
\end{thebibliography}
\end{document}